\documentstyle[prb,aps,twocolumn,epsf]{revtex}
\title{Ground state energy of the Hubbard model: cluster--perturbative results}
\author{Maciej M. Ma\'ska}
\address{Deptartment of Theoretical Physics, Silesian University,\\
Uniwersytecka 4, 40--007 Katowice, Poland \\
e--mail: maciek@risc3.phys.us.edu.pl}
\begin{document}
\maketitle
\renewcommand{\thefootnote}{\fnsymbol{footnote}}
\begin{abstract}
When electron correlations are important it is often necessary to
use numerical methods to solve the Hamiltonian for a finite system
(cluster) ``exactly''. Unfortunately, such methods are restricted to small 
systems. We
propose to combine the ``exact'' numerical diagonalization for small clusters
with the perturbative calculations to take into account the intracluster as 
well as intercluster interactions.
\end{abstract}

\section{Introduction}

Since the discovery of high--$T_c$ superconductivity \cite{htsc}, the
behavior of strongly correlated electronic systems remains a central problem
in contemporary condensed matter physics. In spite of considerable effort 
devoted to the analysis of these systems, it is clear that the necessary
theoretical skills and tools to deal with strongly correlated fermion
systems are lacking. There are no exact solutions except in one dimension 
(e.g., the $t-J$ model is exactly solved by the Bethe--ansatz method for 
$J=2t$ \cite{tj}) and approximate analytic techniques have been known to 
lead to qualitatively incorrect predictions. The fundamental obstacle which 
appears in the analytical approaches is the difficulty in handling the 
strong correlations in a satisfactory way. Moreover, in mean--field 
calculations it is necessary to make {\it a priori} assumptions about 
the ground--state properties.

Therefore, most work on models with strongly correlated electrons has been
done using numerical techniques. Among others, variational calculations 
\cite{var}, various realization of quantum Monte Carlo simulations, \cite{QMC}
and an exact diagonalization of small systems \cite{Dagotto} are used to 
obtain the properties of these models.

Unfortunately, numerical methods also meet some serious problems. The main
difficulty in the quantum Monte Carlo calculations is the famous sign
problem, which reduces the usage of this method at low temperatures and at
the physically interesting densities. The minus sign problem does not arise
in the diagonalization procedures, based on the Lancz\"os method \cite{lancz}
and its modifications \cite{lancz1}, where all 
quantities (static and dynamical) can be computed from the ground state.
Regretfully, the Lancz\"os technique is limited to small clusters by the 
rapid increase of the size of the Hilbert space with the number sites. 
Typically the calculations are performed on $4\times 4$ cluster with 
periodic boundary conditions for one hole, two holes, or an arbitrary number 
of holes.
With respect to the infinite lattice this corresponds to an investigation
of only a small number of points in the Brillouin zone. Therefore, the 
overall shapes of the energy bands cannot be determined precisely, and 
 the influence of the cluster size on the eigenstates of the
Hamiltonian can be important. The differences between results obtained 
for a finite cluster, and its values for an infinite lattice, are known as
{\it finite size corrections}. It is often difficult to extrapolate
the finite cluster data to the thermodynamic limit, and in certain
cases it can lead to erroneous theoretical predictions. 
The corrections often do not decrease monotonically with the 
increase of the number of sites, mainly due to varying cluster 
geometries. Thus the estimation of finite size corrections by 
direct comparison of clusters of different sizes is difficult.
Moreover, often one cannot compare different clusters with the 
same filling, as the number of sites as well as the number of
electrons are integers. 

There are various methods of minimizing the finite size corrections.
The most obvious one, the increase of the size of the cluster, is 
strongly limited by the available time and memory of present--day 
computers. However, there have been some recent attempts to attack this
problem, e.g., the diagonalization of the Hamiltonian in a reduced
Hilbert space. \cite{trunc} Another approach to finite size effects
is based on a specific treatment of boundary conditions. Usually, when 
the hopping term in the Hamiltonian makes a particular jump out of 
the cluster, it is mapped back into the cluster through a translation
without any change of the wave function. However, in order to reduce
the finite size effects, twisted boundary conditions are sometimes 
used, i.e., the phase of the wave function is changed when the electron
hops from one site to another. Then the properties of a larger system
may be found by forming an average over smaller systems with different
boundary boundary conditions \cite{IBC}. In another method the boundary
conditions are randomized by varying the magnitude rather then the
phase of the ``boundary'' hopping.

The aim of the present work is to evaluate the ground state energy of the
Hubbard Hamiltonian by employing yet another approach to the cluster
calculations. Instead of applying some kind of boundary condition,
we propose to mimic the infinite lattice by treating the electron
hopping into or from a cluster as a small perturbation, and carrying out
the summation of the perturbation series. 

\section{Fundamentals of the Cluster--Perturbative (CP) Method}
 
\begin{figure}
\epsfxsize=8.5cm 
\centerline{\epsffile{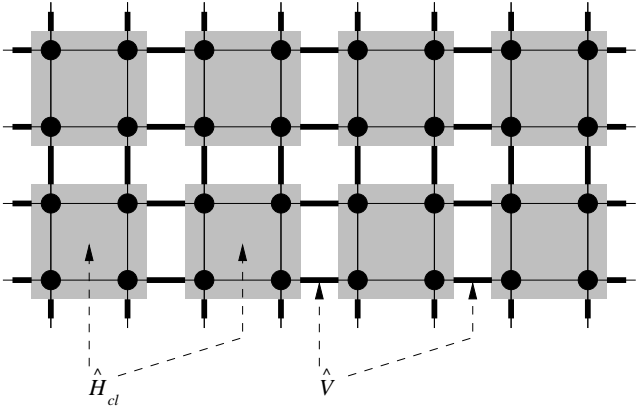}}
\caption{The lattice divided into clusters. $\hat{H}_{cl}$ are the cluster 
Hamiltonians and $\hat{V}$ is the terms describing the hopping between 
 clusters.}
\label{lattice}
\end{figure}

The idea of the present approach is to divide the infinite lattice
into small exactly soluble clusters and consider the transfer between
the clusters as a perturbation (see Fig. \ref{lattice}). 
The Hamiltonian of the system consists of two parts:
\begin{equation}
\hat{H}=\hat{H}_{cl}+\hat{V},
\end{equation}
where
\begin{equation}
\hat{H}_{cl}=\sum_{I=1}^{\cal N} \hat{H}_I
\end{equation}
is the sum of the cluster Hamiltonians, and
\begin{equation}
\hat{V}=\sum_{<IJ>} \hat{V}_{IJ}
\end{equation}
describes the hopping between nearest neighboring clusters. 
${\cal N}$ denotes the number of clusters.

The cluster Hamiltonian $\hat{H}_I\ (I=1,..,{\cal N}) $ operates on the 
$I$th cluster's Hilbert space ${\cal H}_I\ (\hat{H}_I:\ {\cal H}_I 
\rightarrow {\cal H}_I)$ and the Hilbert space of the whole system is 
a direct product of the cluster's Hilbert spaces ${\cal H}={\cal H}_1 
\otimes {\cal H}_2 \otimes\ . . . \otimes {\cal H}_{\cal N}$. Taking the 
size of the clusters relevant for numerical diagonalization, we are 
able to separately solve the cluster Hamiltonians. Now, in order to 
model the infinite lattice, we turn on the intercluster electronic 
transfer. The operator that moves electrons between clusters $I$ and 
$J$ ($I$ and $J$ are neighboring clusters) operates in a direct product of 
the Hilbert spaces of clusters $I$ and $J$:
\begin{equation}
\hat{V}_{IJ}:\ {\cal H}_I \otimes {\cal H}_J \rightarrow {\cal H}_I 
\otimes {\cal H}_J.
\end{equation}
Assuming that the intercluster hopping energy is small compared to the 
distances between levels in the cluster's spectrum, we can perform 
perturbation calculations, where the states of the Hamiltonian 
$\hat{H}_{cl}$ will be the zeroth-order approximation.

Let $\{|\phi^I_i\rangle\},\ i=0,1,...$ denote the set of states of 
$I$th cluster, and $\{\epsilon_i\}$ the corresponding energy levels
$(\epsilon_0 \leq \epsilon_1 \leq \epsilon_2 \leq ... )$. 
Then 
\begin{equation}
|\Phi_{\{i_1,i_2,..,i_{\cal N}\}}\rangle\,=\,|\phi^1_{i_1}\rangle\otimes 
|\phi^2_{i_2}\rangle\otimes ... \otimes |\phi^{\cal N}_{i_{\cal N}}\rangle
\end{equation}
is an eigenstate of $\hat{H}_{cl}$ corresponding to the energy
\begin{equation}
E_{\{i_1,i_2,..,i_{\cal N}\}}\,=\,\epsilon_{i_1}+\epsilon_{i_2}+
..+\epsilon_{i_{\cal N}}.
\end{equation}
The corrections to the ground-state energy $E_0$ in the $n$th 
order of the perturbation $\hat{V}$ are calculated summing the 
Goldstone diagrams
\begin{equation}
\Delta E_n\,=\,\langle\Phi_0|\,\hat{V}\left(\frac{1}{E_0-\hat{H}_{cl}}
\hat{V}\right)^{n-1}|\Phi_0\rangle_{\rm conn},\ \ n=1,2,...\ ,
\label{gold}
\end{equation} 
where 
$|\Phi_0\rangle\,\equiv\,|\phi^1_0\rangle\otimes |\phi^2_0\rangle\otimes ... \otimes 
|\phi^{\cal N}_0\rangle$ is the ground state of the Hamiltonian 
$\hat{H}_{cl}$.

\section{Application of the CP formalism to Two-Dimensional Hubbard model.}

On a simple square lattice,
$\Delta E_n$ vanishes for odd $n$ since the perturbation $\hat{V}$ 
transfers an electron from one cluster to a neighboring one, and
it needs an even number of jumps to return the electron to the
outgoing cluster creating the ground state again.

The lowest-order nonvanishing contribution to the ground-state energy
$\Delta E_2$ is given by:
\begin{equation}
\Delta E_2\,=\,{\sum_{i_1,i_2,..,i_{\cal N}}}'  
\frac{{\bf |}\langle\Phi_0|\hat{V}
|\Phi_{\{i_1,i_2,..,i_{\cal N}\}}\rangle{\bf |}^2}{E_0-E_{\{i_1,i_2,..,
i_{\cal N}\}}},
\end{equation}
where the prime means summing over all states excluding the ground 
state. In the case of simple square lattice the symmetry of the lattice
allows to simplify this expression, leading to
\begin{equation}
\Delta E_2\,=\,2{\cal N}{\sum_{i_1,i_2}}' 
\frac{{\bf\mid}\langle\Phi_{\{0,0\}}|
\hat{V}_{12} |\Phi_{\{i_1,i_2\}}\rangle{\bf\mid}^2}
{2\,\epsilon_0 - \epsilon_{i_1} 
- \epsilon_{i_2}}.
\end{equation}

\begin{figure}
\epsfxsize=8.5cm
\centerline{\epsffile{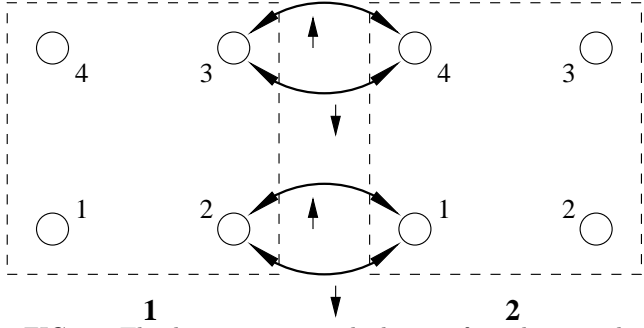}}
\caption{The hopping term which transfers electrons between clusters 1 and 2
($\hat{V}_{12}$).}
\label{v12}
\end{figure}

The operator $\hat{V}_{12}$ is given by
\begin{equation}
\hat{V}_{12} = -t \sum_\sigma\left(
a^\dagger_{3,1,\sigma}a_{4,2,\sigma} + 
a^\dagger_{2,1,\sigma}a_{1,2,\sigma}\right) + {\rm H.c.},
\label{eqv12}
\end{equation}
where we have used two indices for the creation and annihilation 
operators: the first indicates the position of a site
within the cluster, and the second index is the number of the cluster.
In Fig. \ref{v12} the solid lines connecting clusters 1 and 2 represent 
different terms of the operator $\hat{V}_{12}$ [Eq. (\ref{eqv12})].

The matrix element $\langle\Phi_{\{0,0\}}|\hat{V}_{12} 
|\Phi_{\{i_1,i_2\}}\rangle$ 
can be expressed in terms of $\langle\phi_0|a_{l,\sigma}|\phi_i\rangle$ and
 $\langle\phi_0|a^\dagger_{l,\sigma}|\phi_i\rangle$ ($\sigma=\uparrow,\downarrow;\ 
l=1,..,\cal{M}$, where $\cal{M}$ is the number of sites in the cluster), 
where
$a_{l,\sigma}\ (a^\dagger_{l,\sigma})$ creates (annihilates) an electron 
with spin $\sigma$ on the $l$th site in the cluster. 
Thus all the calculations required to evaluate the second-order
correction are performed within Hilbert space of a size
equal to the size of Hilbert space of a single cluster.

\begin{figure}
\epsfxsize=8.5cm
\centerline{\epsffile{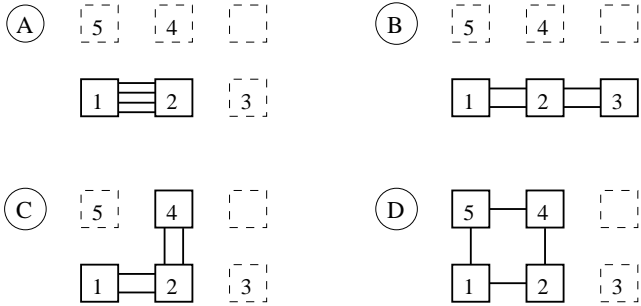}}
\vspace{3mm}
\caption{The diagrams which contribute to the fourth order corrections.}  
\label{diags}
\end{figure}

In the same manner, we are able to take into account the fourth-order
correction $\Delta E_4$. While the computational potential required to 
perform such calculations is much larger (mainly due to a summation over 
a large number of intermediate states), the Hilbert space is still the 
same as in the zeroth order. The general formula for the fourth-order 
correction is too complicated to be presented here. Instead, Fig.
\ref{diags} shows all the diagrams that contribute to $\Delta E_4$ in the 
case of a simple square lattice. 

The lines between clusters $I$ and $J$ represents 
a perturbation $\hat{V}_{IJ}$ that moves an electron between
these clusters (in both directions, from $I$ to $J$ as well as from
$J$ to $I$). 
The interaction in Eq. (\ref{gold}) operates on the ground state, and 
as a result also produces the ground state. Therefore, only diagrams of
the forms of loops contribute to the ground-state energy and, for example, the
following diagram does not appear in the fourth order:
$\Box\!\rightarrow\!\Box\!\rightarrow\!\Box\!\rightarrow\!\Box
\!\rightarrow\!\Box$.
The simplest diagram (A) gives the contribution which can be written 
explicitly as
\begin{eqnarray}
\Delta E^A_4 & = & \Gamma_A \sum_{i_1,i_2} \sum_{j_1,j_2} \sum_{k_1,k_2} 
      \langle\Phi_{\{0,0\}}|\hat{V}_{12}|\Phi_{\{i_1,i_2\}}\rangle
\nonumber \\
 & \times & \langle\Phi_{\{i_1,i_2\}}|\hat{V}_{12}|\Phi_{\{j_1,j_2\}}\rangle 
\nonumber \\
 & \times & \frac{
      \langle\Phi_{\{j_1,j_2\}}|\hat{V}_{12}|\Phi_{\{k_1,k_2\}}\rangle
      \langle\Phi_{\{k_1,k_2\}}|\hat{V}_{12}|\Phi_{\{0,0\}}\rangle}
      {(2\epsilon_0 - \epsilon_{i_1} - \epsilon_{i_2})
       (2\epsilon_0 - \epsilon_{j_1} - \epsilon_{j_2})
       (2\epsilon_0 - \epsilon_{k_1} - \epsilon_{k_2})} . \nonumber \\
\end{eqnarray}
A contribution from each diagram is multiplied by 
the factor which reflects the symmetry of the diagram and the lattice; 
for example there are four diagrams of type A (obtained by rotating 
diagram A around site 1 by $90^\circ$), and each one consists of 
two sites, so that $\Gamma_A=2$. 

Equations for contributions from other diagrams are more complicated,
for example diagram B includes the following 
processes: 
\begin{eqnarray}
1\rightarrow 2,2\rightarrow 3,3 \rightarrow 2,2 \rightarrow 1
,\ \ \ &
1\rightarrow 2,3\rightarrow 2,2 \rightarrow 3,2 \rightarrow 1,
\nonumber \\ 
2\rightarrow 1,2\rightarrow 3,3 \rightarrow 2,1 \rightarrow 2
,\ \ \ &
2\rightarrow 1,3\rightarrow 2,2 \rightarrow 3,1 \rightarrow 2,
\nonumber \\
1\rightarrow 2,2\rightarrow 3,2 \rightarrow 1,3 \rightarrow 2
,\ \ \ &
1\rightarrow 2,3\rightarrow 2,2 \rightarrow 1,2 \rightarrow 3,
\nonumber \\
2\rightarrow 1,2\rightarrow 3,1 \rightarrow 2,3 \rightarrow 2
,\ \ \ &
2\rightarrow 2,3\rightarrow 2,1 \rightarrow 2,2 \rightarrow 3,
\nonumber
\end{eqnarray}
where $I\rightarrow J$ describes the hopping from cluster $I$ to 
cluster $J$.

The summation over all intermediate states is a difficult, the most 
time--consuming task. In order to reduce the computational effort we have 
explicitly exploited various symmetries of the model. For example,
calculating the matrix element 
$\langle\Phi_{\{i_1,i_2\}}|\hat{V}_{12}|\Phi_{\{j_1,j_2\}}\rangle$,
conservation of the number of particles reduces the subspace of states 
$|\Phi_{\{j_1,j_2\}}\rangle$ to $\left({\cal H}_{N(i_1)+1}\otimes
{\cal H}_{N(i_2)-1}\right) 
\:\oplus\:\left({\cal H}_{N(i_1)-1}\otimes{\cal H}_{N(i_2)+1}\right)$ where 
${\cal H}_N$ denotes a subspace of the cluster's Hilbert space with $N$
electrons and $N(k_m)$ is the number of electrons in the $k$th state on 
the $m$th cluster. 

\section{Results and Discussion}
\begin{figure}
\epsfxsize=8.5cm
\centerline{\epsffile{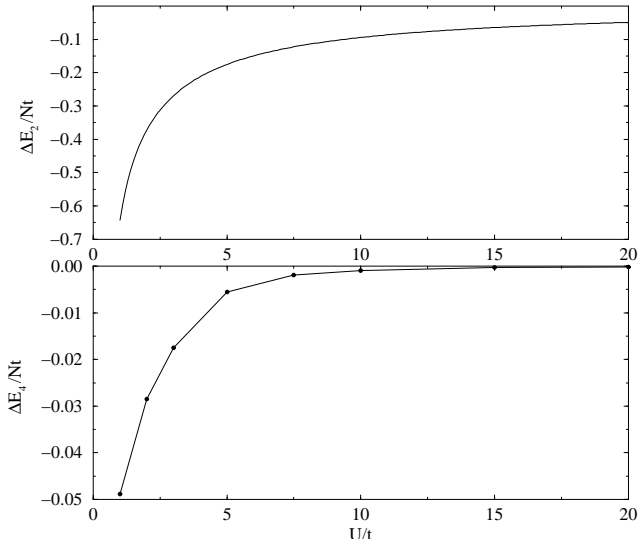}}
\caption{The second and fourth order corrections to the ground state energy.}
\label{corr}
\end{figure}

\begin{figure}
\epsfxsize=8.5cm
\centerline{\epsffile{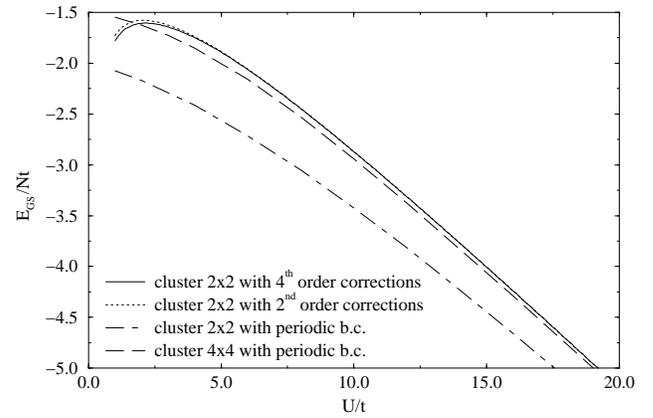}}
\caption{The comparison of ground state energy calculated using the CP
approach with the result of Lancz\"os diagonalization of $4\times 4$
system.} 
\label{gs}
\end{figure}

The second- and fourth-order corrections to the ground-state energy were
calculated for different values of $U/t$ (for $n=1$). The results, presented
in Fig. \ref{corr} and \ref{gs}, are compared with the energies obtained 
by exact diagonalization of the Hamiltonian for $4\times4$ system with periodic
boundary conditions\cite{4x4}.\footnote{In Ref. \cite{4x4} the Hubbard
Hamiltonian is written in particle--hole symmetric form, so the ground
state energies for $n=1$ are shifted by $U/4$ (per site)} 

Generally, apart from the region of $U/t \leq 1$, where the perturbation 
series does not converge, the results from both these approaches are in 
agreement. 
The calculations were performed on IBM RS/6000 workstations, whereas 
diagonalization of $4\times4$ clusters requires much larger computing 
facilities (see, e.g., Ref. \cite{Dagotto}). The advantage of the CP approach,
comparing to the exact diagonalization of larger systems is the lack of
the memory limitations. With the increase of the order only the 
computational time increases, whereas the of size the diagonalized matrices
is constant. Of course, this method, in contradiction to the standard
Lancz\"os approach, does not allow a calculation of the dynamical properties
of a given Hamiltonian. The formalism can be directly applied to systems
described by other than Hubbard Hamiltonians, e.g., the $t-J$ model.

The most attractive application of the CP formalism is the study of
the hole--hole effective interaction.
Performing calculations for system with one and two holes in one cluster,
while all the other clusters are with $n=1$, we can calculate the
two--hole binding energy:
$\Delta = E({\rm two\ holes})+E_0-2\,E({\rm one\ hole})$. 
Work in this direction is in progress.
However, in the case of a doped cluster we have to perform the calculations
for a degenerated spectrum, where the CPU and memory usage is much larger.

This scheme can be directly extended to a study of the nature of ground
states of the undoped and doped systems. 

\end{document}